\documentstyle[11pt]{article}
\begin{document}
\vskip 5pt
\begin{center}
{\Large {\bf Solar neutrino results and Violation of the Equivalence
Principle: An
analysis of the existing data and predictions for SNO}}
\vskip 20pt

\renewcommand{\thefootnote}{\fnsymbol{footnote}}

{\sf Debasish Majumdar $^{a,b,\!\!}$
\footnote{E-mail address: debasish@tnp.saha.ernet.in}},   
{\sf Amitava Raychaudhuri $^{a,\!\!}$
\footnote{E-mail address: amitava@cubmb.ernet.in}}, 
and 
{\sf Arunansu Sil $^{a,c,\!\!}$
\footnote{E-mail address: arun@cubmb.ernet.in}}  
\vskip 10pt  
$^a${\it Department of Physics, University of Calcutta, 92
Acharya Prafulla Chandra Road, \\ Calcutta 700009, India.}\\ 
$^b${\it Saha Institute of Nuclear Physics,\\ 
1/AF Bidhannagar, Calcutta 700064, India. }\\
$^c${\it Department of Physics, Jadavpur University,
Calcutta 700032, India.}\\
\vskip 15pt
{\bf Abstract}
\end{center}

{\small 
Violation of the Equivalence Principle (VEP) can lead to neutrino
oscillation through the non-diagonal coupling of neutrino flavor
eigenstates with the gravitational field. The neutrino energy
dependence of this oscillation probability is different from that of
the usual mass-mixing neutrino oscillations. In this work we explore,
in detail, the viability of the VEP hypothesis as a solution to the
solar neutrino problem in a two generation scenario with both the
active and sterile neutrino alternatives, choosing these states to be
massless.  To obtain the best-fit values of the oscillation parameters
we perform a $\chi^2$ analysis for the total rates of solar neutrinos
seen at the Chlorine (Homestake), Gallium (Gallex and SAGE),
Kamiokande, and SuperKamiokande (SK) experiments. We find that
the goodness of these fits is never satisfactory. It markedly
improves, especially for VEP transformation to sterile neutrinos,
if the Chlorine result is excluded from the analysis. The
1117-day SK data for recoil electron spectrum are also examined
for signals of VEP oscillations. For these fits, we consider
variations of the Standard Solar Model by allowing the absolute
normalizations of the $^8$B and $hep$ neutrinos to vary. Here the
fits are quite good but the best fit values of the parameters are
rather different from those from the total rates fits. A combined
fit to the total rates and recoil electron spectrum data is also
performed. We present the 90\% confidence limit contours for all
the three analyses mentioned above.  The best-fit parameters
obtained from the recoil electron spectrum and the combined
analysis of rate and spectrum are used to predict the charge
current and scattering electron spectrum at SNO.  }

\vskip 20pt

\begin{center}
PACS NO. 26.65.+t, 14.60.Lm, 14.60.Pq, 04.80.Cc
\end{center}

\newpage
\renewcommand{\thesection}{\Roman{section}}
\renewcommand{\thefootnote}{\arabic{footnote}}
\setcounter{footnote}{0}
\section{Introduction}

Oscillation of neutrinos from one flavor to another is currently the
favored solution to the solar neutrino problem \cite{sk1117,sksolar,
solar}. This proposition has been strengthened by the recent
SuperKamiokande (SK) atmospheric neutrino data \cite{atm} which also
support the existence of neutrino oscillations.  The usual formulation
of neutrino oscillations rests on two essential properties; namely, (a)
the neutrinos are massive and are further not mass degenerate, and (b)
the flavor eigenstates ({\em i.e.,} $\nu_e$, $\nu_\mu$, and $\nu_\tau$)
are not themselves the eigenstates of mass but rather linear
superpositions of the latter.  Distinct from the simplest vacuum
oscillations, neutrino flavor conversion can also be induced by
the passage of neutrinos in matter (in this case solar matter)
due to the difference in the strength of weak interactions of
neutrinos of different flavor with ambient electrons. This is the
Mikheyev-Smirnov-Wolfenstein (MSW) \cite{msw} effect.  Vacuum
oscillations and the MSW effect are widely considerd to be strong
candidates for the solution of the solar neutrino problem. Much
work has already been done to analyze the available solar
neutrino data in terms of these alternate possibilities and
substantial effort is still being devoted to obtain the
parameters that best fit the data \cite
{mswbks,valle,bkssno,fl99,gg4,gmr}.

Though the vacuum oscillation and MSW solutions relying on
massive neutrinos (we refer to these collectively as
`mass-mixing' solutions henceforth) are by far the most popular
scenarios for addressing the solar and atmospheric neutrino data,
oscillations can also originate from other sources. One such
possibility presents itself if violation of the weak equivalence
principle (VEP) occurs and the flavor eigenstates are not
identical to the states that couple to gravity. The principle of
equivalence is a cornerstone of Einstein's general theory of
relativity. Normally such a premise would be considered
sacrosanct, but so little has been experimentally tested for
neutrinos that it may not be unreasonable to keep an open mind
and check the validity of this principle for them. If this
principle is indeed violated then, as a consequence, the coupling
of neutrinos to the gravitational field is nonuniversal. Under
this circumstance, if the flavor eigenstates are linear
superpositions of the gravitational eigenstates,  VEP induced
oscillations of neutrinos take place \cite{gasp}. This does not
require neutrinos to carry a non-zero mass. The important
difference between this approach and the mass-mixing solution is
manifested in the energy dependence of the survival probability.
For a two-neutrino picture, the general expression for the
survival probability for an initial $\nu_e$ after propagation
through a distance $L$ is given by:
\begin{equation}
P_{ee}(E_\nu,L) = 1 - \sin^22\theta \sin^2
\left( \frac {\pi L}{\lambda}\right),
\label{eq:sprob}
\end{equation}
where $\theta$ is the rotation angle relating  the gravitational
eigenstate basis ($\nu_1$, $\nu_2$) to the flavor basis
($\nu_e$, $\nu_x$, $x = \mu$, $\tau$, or a sterile state).
\begin{equation}
\nu_e = \nu_1 \cos \theta + \nu_2 \sin \theta; \;\;\;\;
\nu_x = - \nu_1 \sin \theta + \nu_2 \cos \theta
\label{eq:mix}
\end{equation}
and $\lambda$ is the oscillation length, which for the VEP induced
oscillation is:
\begin{equation}
\lambda = \frac {2 \pi}{E_\nu \phi \Delta f},
\label{eq:veplam}
\end{equation}
where $E_\nu$ is the neutrino energy, $\phi$
the gravitational potential, and $\Delta f = f_1 - f_2$ is a
measure of the violation of equivalence principle, $f_i$ being
the coupling strength of the gravitational eigenstates. In
contrast, for mass-mixing vacuum oscillations $\lambda = {(4\pi
E_\nu)}/{\Delta m^2}$, where $\Delta m^2$ is the mass square
difference between two neutrino species.  Thus, for the latter case
$\lambda \propto E_\nu$ while for VEP $\lambda \propto 1/E_\nu$.
Due to the different energy dependences of the survival
probability in the mass-mixing and the VEP alternatives, their
predictions can be quite different. The phenomenological
consequences of VEP-driven neutrino oscillations has attracted
attention over the past decade \cite{vep}.

A completely unrelated situation which also leads to neutrino
oscillations with $\lambda$ $\propto 1/E_\nu$ is a recently
proposed picture of violation of special relativity (VSR)
\cite{coleman}.  If special relativity is violated, the maximum
attainable speed of a particle {\em in vacuo} need not
universally be the speed of light $c$. In particular, if the
maximum possible velocities of two types of neutrinos be $v_1$
and $v_2$ and these {\em velocity eigenstate} neutrinos be
related to the $\nu_e$ and $\nu_x$ through a mixing angle
$\theta$ (see Eq.(\ref{eq:mix})) then the survival probability of
a $\nu_e$ takes the same form as Eq.(\ref{eq:sprob}). In this
case the expression for $\lambda$ is:
\begin{equation}
\lambda = \frac {2 \pi}{E_\nu \Delta v}, 
\label{eq:vsrlam}
\end{equation}
where $\Delta v$ is the velocity difference for the neutrinos
$\nu_1$ and $\nu_2$. Comparing Eqs. (\ref{eq:vsrlam}) and
(\ref{eq:veplam}) one finds that the energy dependence of the
oscillation length is identical in the two cases\footnote{It has
been shown that inclusion of CPT-violating interactions in
addition to Violation of Special Relativity can lead to more
general energy dependences involving $1/E_\nu$, $E_\nu$, and
constant terms \cite{CPT}}  and the role of
$\Delta v$ in the VSR case is the same as that of $\phi \Delta f$
in the VEP formalism. Here, we use the terminology of the
VEP mechanism but the results can be taken over {\em mutatis
mutandis} to the VSR situation.

In this work we make a detailed examination of the  VEP scenario
in the light of the solar neutrino data. We consider the two
possibilities of oscillation of the electron neutrino to (a)
another active neutrino ($\nu_\mu$ or $\nu_\tau$), and (b) to a
sterile ({\em i.e.}, no weak interactions) neutrino. The Chlorine
and Gallium experiments use radiochemical neutrino detection
techniques and do not distinguish between oscillation of the
$\nu_e$ to an active neutrino or to a sterile one.  The
Kamiokande and SuperKamiokande experiments, on the other hand,
are sensitive to other active neutrinos {\em via} the smaller
neutral current contribution.  Hence, the latter discriminates
between the two alternatives of oscillation to active or sterile
neutrinos.

In section II, we present a brief summary of the ingredients that go
into our analysis. 

In the next section we consider the total rates of solar neutrinos
observed by the Chlorine, Gallex, SAGE, Kamiokande, and SuperKamiokande
(1117-day data) experiments\footnote{In the following, we refer to this
data set as the `total rates'.} \cite{solar,sk1117,sksolar}.  We obtain
the best-fit values of the VEP parameters and find that the
goodness-of-fit\footnote{The goodness of fit gives the probability for
the actual $\chi^2$ to exceed $\chi^2_{min}$.} (g.o.f) is never high.
We trace the origin of this to the difficulty within the VEP mechanism
of simultaneously satisfying the Chlorine detector results and the very
precise measurements of SK.  We examine the effect of excluding either
the Chlorine or the Kamiokande result in the fits or of using the
average of the two Gallium results rather than their individual
measurements and find that the best-fit values are markedly different
only in one situation, the exclusion of the Chlorine data from the
analysis, when the quality of the fit is significantly improved.

In section IV we turn to the 1117-day recoil electron energy
spectrum from SK \cite{sk1117} and test the ability of the VEP
model to account for the observations. In this case the quality
of the fit is rather good. We examine a variant  of the Standard Solar
Model (SSM) in
which the absolute normalization of the $^8$B neutrinos ($X_B$)
is allowed to vary. In our analysis we have included the
contribution from $hep$ neutrinos and we have also examined the
situation when the normalization of this flux ($X_h$) is
different from the SSM value of unity. 

Section V deals with the simultaneous fitting of the total rates
data and the SK 1117-day electron energy spectrum results. We
find that good fits can be obtained, in this case, only if $X_B$
and $X_h$ are allowed to assume values different from the SSM
stipulations.

In section VI we turn to the expectations for SNO. We use the
best fit values obtained from the above analyses to check how the
predictions for the charged current deuteron disintegration and
electron scattering are affected by VEP.

We end in section VII with the conclusions. Similar work
analysing solar neutrino data using the VEP formalism have been
performed in the past \cite{bahkra,minak} and more recently in
\cite{kuo,gago,casini}.  We compare our findings with these
results.

\section{Solar neutrinos and VEP}

The sun serves as a good  neutrino factory. The fusion reactions
that generate solar energy also produce neutrinos which are
usually denoted by the different reactions ({\em e.g.}, $pp$, $Be$,
$B$, $hep$, {\em etc.}) from which they originate. The shape of
the neutrino energy spectrum from any reaction is precisely known
from weak interaction theory while the absolute normalization of
the spectrum depends on the solar parameters like core
temperature, opacity, {\em etc.} The $pp$ neutrinos are the most
copious but they are also of the lowest energy. Only the Gallium
experiments (Gallex and SAGE) are sensitive to $pp$ neutrinos and
receive their dominant contributions from this process, in
addition to smaller ones from the other reactions. At the other
extreme, the Kamiokande and SuperKamiokande experiments are based
on the water \u{C}erenkov technique and can detect neutrinos of
energy higher than about 5 MeV. Only the $^8$B neutrinos (and to
a small extent the $hep$ neutrinos) contribute at these energies.
The Chlorine experiment has a threshold of 0.8 MeV and mainly
counts $^7$Be and $^8$B neutrinos. Thus the different detectors
have probed different regions of the solar neutrino spectrum.
Assuming that oscillations do occur during the passage of the
neutrinos from their point of origin to the detector, the
dependence of this phenomenon on the neutrino energy can be
tested by the complementary information from the different
experiments. From these results one should be in a position to
check which of the vacuum, MSW, or VEP oscillations is preferred
by the results.  The SuperKamiokande collaboration has also
presented the observed energy spectrum of the electron scattered
by the incident neutrino. This spectrum is a direct probe of the
energy dependence of neutrino oscillations.

In this work we use the BP98 calculation \cite{bp98} of the solar
neutrino spectrum with INT normalization as the SSM reference. In
addition to this, we explore the possibility of the absolute
normalizations of the $^8$B- and $hep$-neutrinos spectra being
different from their BP98 SSM predictions. If $X_B$ and $X_h$ denote
the factors by which the absolute normalizations are multiplied, we use
the data to find the best-fit values for these. In particular, we find
that the fit to the SK recoil electron spectrum data is improved in a
noteworthy manner when $X_{h}$ is permitted to assume large
values\footnote{See, however, remarks in the penultimate
paragraph of section VII on the fits to the separate day and
night SK electron spectra measurements.}.

The weak equivalence principle requires the coupling of particles
to the ambient gravitational potential $\phi$ to be uniform, {\em
i.e.,} of the form $f \phi E$, where  $E$ is the particle energy,
and $f$ a universal coupling constant. If the latter varies from
one neutrino species to another then that would constitute a
violation of the equivalence principle. If $f_1 \neq f_2$ in a
two-neutrino framework, then these states define a basis in the
two-dimensional space which, in general, could be different from
the flavor basis. The effect due to a small splitting $\Delta f$
will manifest itself in the form of flavor oscillations, the
wavelength going to infinity as $\Delta f$ tends to zero.  We
follow the prevalent practice of choosing the gravitational
potential, $\phi$, to be a constant over the neutrino path. This
is the case if the potential due to the Great Attractor
\cite{dress} dominates over that due to the sun and other
heavenly bodies in our neighborhood. In such an event, writing
$\Delta F = \phi \Delta f/2$, the expression for the oscillation
wavelength, Eq. (\ref{eq:veplam}) becomes
\begin{equation}
\lambda = \frac{\pi}{E_\nu \Delta F} = \frac{6.20 \times 10^{-13}
{\rm m}}{\Delta F}
\left(\frac{1 {\rm MeV}}{E_\nu}\right).
\label{eq:lammag}
\end{equation}
We choose the neutrinos to be massless.

\section{Observed rates and VEP oscillation of neutrinos}

The data used for the $\chi^2$ analysis of total rates are given
in Table 1.  They are from the Chlorine experiment at Homestake,
the two Gallium experiments, Gallex and SAGE, the water
\u{C}erenkov detector Kamiokande, \cite{solar}, and the latest
1117-day data from Super-Kamiokande \cite{sk1117}.

\begin{center}
\begin{tabular}{|c|c|c|c|c|c|}
\hline
Experiment & Chlorine & \multicolumn{2}{|c|}{Gallium}&
Kamiokande&Super\\ \cline{3-4}
& & Gallex & SAGE & &Kamiokande \\ \hline
$\frac{\rm Observed \;\;
Rate}{\rm BP98 \;\; Prediction}$ & $0.33 \pm 0.029$ & $0.60 \pm
0.06$ & $0.52 \pm
0.06$ & $0.54 \pm 0.07$ & $0.465 \pm 0.015$ \\ \hline
\end{tabular}
\end{center}
 
\begin{description}
\item{\small \sf Table 1:} {\small\sf The ratio of the observed
solar neutrino rates to the corresponding BP98 SSM predictions
used in this analysis. The results are from Refs. \cite {solar}
and \cite {sk1117}.}
\end{description}

The definition of $\chi^2$ used for this analysis is:
\begin{equation}
\chi^2 =
\sum_{i,j=1,5} \left(F_i^{th} -
F_i^{exp}\right)
(\sigma_{ij}^{-2}) \left(F_j^{th} - F_j^{exp}\right).
\end{equation}
Here $F_{i}^{\xi}= {T_i^{\xi}}/{T_{i}^{BP98}}$ where
$\xi$ is $th$ (for the theoretical prediction) or $exp$ (for
the experimental value) and $T_i$ is the total rate in the $i$-th
experiment.  $F_{i}^{exp}$ is taken from Table 1.  The error
matrix $\sigma_{ij}$ contains the experimental errors, the
theoretical errors and their correlations. The theoretical errors
have contributions which originate from uncertainties in the
detector cross-sections as well as from astrophysics \cite{flap}.
The off-diagonal elements in the error matrix come through the
latter.  In estimating the astrophysical contribution, the
uncertainties in the spectrum due to the input parameters are
taken from \cite{bp98}.

In the presence of neutrino conversions, the detection rate on
earth for the radiochemical Chlorine and Gallium experiments is
predicted to be:
\begin{equation}
T_i^{th} =
\sum_\alpha \int_{E_{th}} X_\alpha \phi _\alpha(E_\nu) \sigma_i
(E_\nu) <P_{ee}(E_\nu)>_\alpha dE_\nu,
\end{equation}
where $\sigma_i (E_\nu)$ is the neutrino capture cross-section
for the $i$-th detector \cite{jnbhome} and $E_{th}$ the neutrino
threshold energy for detection. $\phi_{\alpha}(E_\nu)$ stands for
the neutrino spectrum for the $\alpha$-th source \cite{jnbhome}
and $X_\alpha$ is an overall normalization factor for this
spectrum such that $X_\alpha = 1$ corresponds to the SSM.  The
sum is over all the individual neutrino sources.
$<P_{ee}(E_\nu)>_\alpha$ is the neutrino survival probability for
the $\alpha$-th source averaged over the distribution of neutrino
production regions in the sun,
\begin{equation}
<P_{ee}(E_\nu)>_\alpha = \int dr ~P_{ee}(E_\nu,r)
\Phi_{\alpha}(r).
\end{equation}
$\Phi_{\alpha}(r)$ is a normalized
function  which gives the probability of the $\alpha$-th
reaction occuring at a distance $r$ from the center of the sun and
$P_{ee}(E_\nu,r)$ is obtained by averaging the
survival probability over a year taking
the eccenticrity of the earth's orbit into account, {\em i.e.,}
\begin{equation}
P_{ee} (E_\nu,r) = \frac{1}{T} \int_0^T dt \left[  1 -
\sin^{2}{2\theta}\sin^2\left\{ {\frac{\pi
R(t)}{\lambda}}\left(1-\frac{r}{R(t)}\right)\right\} \right],
\end{equation}  
where  $R(t)$ is the sun-earth distance given by,
\begin{equation}
R(t) = R_{0} \left[ 1 - \epsilon \cos(2 \pi \frac{t}{T})\right].
\end{equation}
Here, $R_{0} = 1.49 \times 10^{11}$ m is the mean Sun-Earth
distance and $\epsilon = 0.0167$ is the ellipticity of the
earth's orbit. $t$ is the time of the year at which the solar
neutrino flux is measured and $T$ is 1 year.

The theoretical prediction according to the BP98 Standard Solar
Model, $T_{i}^{BP98}$, is obtained by setting the survival
probability as 1 in the above.

For the water \u{C}erenkov detectors Kamiokande and
SuperKamiokande, in the case of oscillation to active neutrinos
one has to take into account the contributions to the signal from
the produced $\nu_\mu$ or $\nu_\tau$ {\em via} the neutral
current interactions,
\begin{eqnarray}
T_{K,SK}^{th} & = &
\sum_\alpha\int_{E_{\nu_{min}}}^{E_{\nu_{max}}} dE_\nu
\int_{E_{T_{min}}}^{E_{T_{max}}}
dE_{T} \int_{E_{A_{th}}} dE_{A} \; \rho(E_{A}, E_{T})
\;X_\alpha \;\phi _\alpha(E_\nu) \nonumber \\ 
&  & \left[<P_{ee}(E_\nu)>_\alpha
\frac{d\sigma_{\nu_e}}{dE_{T}} + <P_{e\mu}(E_\nu)>_\alpha
\frac{d\sigma_{\nu_\mu}}{dE_{T}}\right].
\label{rkam}
\end{eqnarray}
The second term in the bracket is absent if oscillation to sterile
neutrinos is under consideration.  $E_{T}$ and $E_{A}$ denote the true
and apparent electron energies respectively.  $E_{T_{min}}$ and
$E_{T_{max}}$ are determined by kinematics.  $\rho (E_{A}, E_{T})$ is
the energy resolution function for which we use the expression given in
\cite{blk}.  $E_{A_{th}}$ is 7.5 (5.5) MeV for the calculation of the
total rate at Kamiokande (SuperKamiokande).  The differential
cross-section for the production of an electron with true relativistic
energy $E_{T}$, $\frac{d\sigma}{dE_{T}}$, is obtained from standard
electroweak theory.

Now we discuss the results obtained from a $\chi^2$ minimization
analysis of the data within the VEP picture.  For these fits we
have set $X_\alpha = 1$ for all $\alpha$; {\em i.e.,} we take the
normalizations of the solar spectra at their SSM values.  The
best-fit parameters, $\chi^2_{min}$/(degree of freedom), and the
goodness of fit  values are presented in Table 2. Both possibilities of
oscillation of the $\nu_e$ to an active or a sterile neutrino have been
considered.  To gauge the impact of the different experiments on the
best-fit values of the VEP parameters, we have first fitted all the
five data given in Table 1. We have then repeated the procedure
excluding the Chlorine or the Kamiokande results. We also examine the
effect of using the average of the result of the two Gallium
measurements ($0.57 \pm 0.054$) rather than the two individual ones.
It is seen from Table 2 that in the different alternatives the best-fit
values of the parameters are all rather close excepting for the case
where the Chlorine result is excluded from the analysis. The fit
improves significantly in the latter case; but even here the goodness
of fit for active (sterile) neutrinos is still just 36\% (58.5\%).  It
is seen that maximal mixing is preferred and the VEP oscillation length
is (see Eq.  (\ref{eq:lammag})) $\sim 3.4 \times 10^{11}$ m, comparable
with the earth-sun distance. This is reminiscent of the mass-mixing
vacuum oscillation solution to the solar neutrino problem.

\begin{center}
\begin{tabular}{|c|c|c|c|c|c|c|}
\hline
Neutrino&Set&Fitted & $\sin^2 2\theta$ & $\Delta F$ &
$\chi^2_{min}$/d.o.f&g.o.f\\
Type&&Experiments&&(10$^{-24})$ && (\%)\\
\hline
&1a&Cl, Gallex, SAGE, K, SK & 1.0 & 1.80 & 4.54/3&20.84 \\
\cline{2-7}
&1b&Gallex, SAGE, K, SK & 0.85 & 4.57 & 2.04/2&36.05 \\
\cline{2-7}
Active&1c&Cl, Gallex, SAGE, SK & 1.0 & 1.80 & 3.50/2&17.41
\\ \cline{2-7} 
&1d&Cl, (Ga)$_{\rm av}$, K, SK & 1.0 & 1.80 & 3.72/2&15.59 \\
\cline{2-7}
&1e&Cl, (Ga)$_{\rm av}$, SK & 1.0 & 1.80 & 2.67/1&10.20 \\
\hline
\hline
&2a&Cl, Gallex, SAGE, K, SK & 1.0 & 1.84 & 5.89/3&11.73 \\
\cline{2-7}
&2b&Gallex, SAGE, K, SK & 0.84 & 4.53 & 1.07/2&58.51 \\
\cline{2-7}
Sterile&2c&Cl, Gallex, SAGE, SK & 1.0 & 1.84 & 4.58/2&10.12 \\
\cline{2-7}
&2d&Cl, (Ga)$_{\rm av}$, K, SK & 1.0 & 1.84 & 5.08/2&7.91 \\
\cline{2-7}
&2e&Cl, (Ga)$_{\rm av}$, SK & 1.0 & 1.84 & 3.77/1&5.23 \\
\hline
\end{tabular}
\end{center}

\begin{description}
\item{\small \sf Table 2:} {\small \sf The best-fit values of the
parameters, $\sin ^22\theta$, $\Delta F$, $\chi^2_{min}$, and the
g.o.f.  for fits to the total rates of the different
experiments.}
\end{description}

As is seen from Table 2, the goodness of fits are rather poor for
all the above cases. In order to trace the origin of this result,
we present in Table 3 the total rates for the different
experiments obtained using the best-fit values of the VEP
parameters presented in Table 2.  These should be compared with
the experimental data in Table 1. Several points are noteworthy.
The rates for the different experiments are not sensitive to the
small changes in the input VEP parameters; the numbers for the
Gallium experiments and Kamiokande are always within 1$\sigma$ of
the experimental value but for Chlorine the deviation is $3
\sigma$. The fit to SuperKamiokande is always bad (8$\sigma$)
irrespective of which experiments are excluded from the analysis
excepting for the singular case where the Chlorine rate is left
out when a 2$\sigma$ fit is obtained.  This is a reflection of
the very precise nature of the present SK data and the inability
of the VEP mechanism to simultaneously reproduce the varying
degrees of suppression seen in experiments with different energy
thresholds. In particular, the Chlorine result is seen to be
especially problematic in this respect. It can be surmised from
the Chlorine-excluded analyses -- Table 3 (sets 1b, 2b) -- that a
simultaneous good fit to all data would be obtained if the
suppression in the Chlorine experiment had been {\em less}  than
that in the Gallium experiments (lower threshold) as well as in
the water \u{C}erenkov ones (higher threshold), indicating
that for these fit-values of the VEP parameters the
$^7$Be neutrinos are less suppressed than the other solar
neutrinos.

\begin{center}
\begin{tabular}{|c|c|c|c|c|c|c|}
\hline
Neutrino&Set&Fitted & Cl & Ga &
K&SK\\ 
Type&&Experiments& &  & & \\ \hline
&1a&Cl, Gallex, SAGE, K, SK & 0.436 & 0.572
&0.600 & 0.600 \\ \cline{2-7}
&1b&Gallex, SAGE, K, SK & 0.655$^*$ & 0.549
& 0.508 & 0.497 \\ \cline{2-7}
Active&1c&Cl, Gallex, SAGE, SK & 0.436 & 0.572
& 0.600$^*$ & 0.600 \\ \cline{2-7}
&1d&Cl, (Ga)$_{\rm av}$, K, SK & 0.436 & 0.574
&0.600 & 0.600 \\ \cline{2-7}
&1e&Cl, (Ga)$_{\rm av}$, SK & 0.436 & 0.574
& 0.600$^*$ & 0.600 \\ \hline
\hline
&2a&Cl, Gallex, SAGE, K, SK & 0.435 & 0.565
&0.525 & 0.525 \\ \cline{2-7}
&2b&Gallex, SAGE, K, SK & 0.657$^*$ & 0.555
&0.512 & 0.468 \\ \cline{2-7}
Sterile&2c&Cl, Gallex, SAGE, SK & 0.435 & 0.565
& 0.525$^*$ & 0.525 \\ \cline{2-7}
&2d&Cl, (Ga)$_{\rm av}$, K, SK & 0.435 & 0.565
&0.525 & 0.525 \\ \cline{2-7}
&2e&Cl, (Ga)$_{\rm av}$, SK & 0.435 & 0.565
& 0.525$^*$ & 0.525 \\ \hline
\end{tabular}
\end{center}

\begin{description}
\item{\small \sf Table 3:} {\small \sf The total rate predictions
for the different experiments obtained by using the best-fit
values of the VEP parameters presented in Table 2. Predictions
for experiments not included in the fit are marked with an
asterisk.}
\end{description}

We show in Fig. 1, the allowed region for the parameters
$\sin^22\theta$ and $\Delta F$ at 90\% C.L. for the (1a) active
and (1b) sterile cases obtained by fitting the total rates of all
five experiments, {\em i.e.}, Chlorine, Gallex, SAGE, Kamiokande,
and SuperKamiokande. The confidence level has been fixed with
respect to the global minimum. The best fit points have been
indicated.  It is seen from Fig. 1 that the nature of the allowed
regions for the active and sterile cases are roughly similar, an
observation made by Gago {\em et al.} \cite{gago}. The sterile
alternative is more restrictive.

\section{SuperKamiokande Recoil electron energy spectrum and VEP}

We now turn to a fit to the recoil electron energy spectrum as
seen at SuperKamiokande.  We use the 1117-day data for this
analysis. The SK results have been presented in the form of
number of events in 17 electron recoil energy bins of width 0.5
MeV in the range 5.5 MeV to 14 MeV and an 18th bin which covers
the events in the range 14 to 20 MeV \cite{sk1117,smy}.

With the SuperKamiokande threshold in the 5 MeV range, only the
$^8$B and $hep$ solar neutrinos contribute to the signal. In
addition to the SSM, for this analysis we have examined models in
which the normalizations of both the $^8$B and $hep$ neutrino
flux ($X_B$ and $X_h$, respectively -- normalized to unity for
the SSM) are allowed to vary arbitrarily.  In this case, $\chi
^2$ is defined as
\begin{equation}
\chi^2 =
\sum_{i,j=1,18} \left(R_i^{th} -
R_i^{exp}\right)
(\sigma_{ij}^{-2})_{sp} \left(R_j^{th} - R_j^{exp}\right).
\end{equation}
$R_i^{\xi} = S^{\xi}_i/S_i^{BP98}$ with $\xi$ being $th$ or $exp$ as before
and $S_i$ standing for the number of events in the $i$-th energy bin.
The theoretical prediction is given by eq. (\ref{rkam}) but the
integration over the apparent ({\em i.e.}, measured) energy will now be
over each bin.  The error matrix $\sigma_{ij}$ used by us is \cite{valle}
\begin{equation}
(\sigma_{ij}^2)_{sp} = \delta_{ij}(\sigma^2_{i,stat} +
\sigma^2_{i,uncorr}) + \sigma_{i ,exp} \sigma_{j,exp} +
\sigma_{i,cal} \sigma_{j,cal},
\end{equation}
where we have included the  statistical error, the uncorrelated
systematic errors and the energy-bin-correlated experimental
errors \cite{skspec} as well as those from the calculation of the
shape of the expected spectrum \cite{shape}. Since we allow the
normalizations of the $^{8}$B and $hep$ fluxes to vary, we do not
include their astrophysical uncertainties separately. The results
are presented in Table 4.

\begin{center}
\begin{tabular}{|c|c|c|c|c|c|c|c|}
\hline
Neutrino&Set&$\sin^2 2\theta$ & $\Delta F$ & $X_B$ & $X_h$ &
$\chi^2_{min}$/d.o.f. & g.o.f \\
Type&&& $(10^{-24})$ & & & & (\%)\\
\hline
&3a&1.0 & $1.97\times10^2$ & 0.79 & 15.07 & 8.83/14&84.2 \\
\cline{2-8}
&3b&0.38 & 0.22 & 0.65 & 1.0  & 10.04/15&81.7 \\
&&&&& (fixed) && \\ \cline{2-8}
Active&3c&0.69 & 0.23 & 1.0 & -2.20 & 11.59/15&70.97 \\
&&&&(fixed) &&& \\ \cline{2-8}
&3d&0.68 & 0.23 & 1.0 & 1.0 & 11.66/16&76.7 \\
&&&&(fixed) & (fixed) && \\ \hline
\hline
&4a&1.0 & 39.33 & 1.03 & 11.52 & 8.78/14&84.48 \\ \cline{2-8}
&4b&0.34 & 0.22 & 0.67 & 1.0 & 10.01/15&81.9 \\
&&&&& (fixed) && \\ \cline{2-8}
Sterile&4c&0.98 & 39.33 & 1.0 & 11.94 & 8.81/15&88.71 \\
&&&&(fixed) &&& \\ \cline{2-8}
&4d&0.58 & 0.23  & 1.0 & 1.0 & 11.34/16&78.8 \\
&&&&(fixed) & (fixed) && \\
\hline
\end{tabular}
\end{center}

\begin{description}
\item{\small \sf Table 4:} {\small \sf The best-fit values of the
parameters, $\sin^22\theta$, $\Delta F$, $X_B$, $X_h$, $\chi^2_{min}$,
and the g.o.f. for fits to the scattered electron spectrum at SK.}
\end{description}

The large values of $X_h$ obtained from the fits with the best
g.o.f. (3a and 4b) indicate that an increased number of high
energy $hep$ neutrinos yields a better fit to the highest energy
bin of the spectrum observed at SK. It may be noteworthy that a
rise in the observed electron energy spectrum at the high energy
end seen in the earlier 825-day data has become less prominent in
the latest 1117-day sample. A further softening will bring $X_h$
closer to the SSM prediction\footnote{See the remarks on fits to
the SK separate day and night spectra measurements in section VII}.

\begin{center}
\begin{tabular}{|c|c|c|c|c|c|c|c|c|c|c|}
\hline
Neutrino&Set&$\sin^2 2\theta$ & $\Delta F$ & $X_B$ & $X_h$ & Cl &
Ga & K & SK \\ Type&&& $(10^{-24})$ & & & & & & \\
\hline
&3a&1.0 & 1.97 & 0.79 & 15.07 & 0.412 & 0.463 & 0.471 & 0.466 \\
&&&$\times 10^2$& &&& &&\\ \cline{2-10}
&3b&0.38 & 0.22 & 0.65 & 1.0 & 
0.585 & 0.947 & 0.481 & 0.488 \\
&&&&& (fixed) &&& &\\
\cline{2-10}
Active&3c&0.69 & 0.23 & 1.0 & -2.20 &
0.560 & 0.944 & 0.497 & 0.503 \\
&&&&(fixed) &&&& &\\
\cline{2-10}
&3d&0.68 & 0.23 & 1.0 & 1.0 & 0.562 &
0.944 & 0.507 & 0.513 \\
&&&&(fixed) & (fixed) &&& &\\
\hline
\hline
&4a&1.0 & 39.33 & 1.03 & 11.52 & 0.562 & 0.538 & 0.479 & 0.477 \\
\cline{2-10} &4b&0.34 & 0.22 & 0.67 & 1.0 & 0.612 & 0.951 & 0.485 &
0.492 \\ &&&&& (fixed) &&& &\\
\cline{2-10}
Sterile&4c&0.98 & 39.33 & 1.0 & 11.94 &
0.563 & 0.548 & 0.477 & 0.476 \\
&&&&(fixed) &&&& &\\
\cline{2-10}
&4d&0.58 & 0.23 & 1.0 & 1.0 & 0.630 &
0.953 & 0.507 & 0.509 \\
&&&&(fixed) & (fixed) &&& &\\
\hline
\end{tabular}
\end{center}

\begin{description}
\item{\small \sf Table 5:} {\small \sf The calculated values of
the rates for the different experiments using the best-fit values of the
parameters, $\sin^22\theta$, $\Delta F$, $X_B$, and $X_h$, 
from fits to the scattered electron spectrum at SK.}
\end{description}

It may be of interest to check how the best-fit values of the VEP
parameters fare when confronted with the total rates data. For
this purpose, we use the four sets of best-fit values of
parameters from Table 4 and use them to compute the total rates
from the different experiments. These results are shown in Table
5. The latter may be compared to the values obtained using fits to
the total rates themselves (see Table 3). Using Table 1, it
is seen that the predicted rate for SK is in rather close
agreement with the measured value, as is to be expected, but for
the Chlorine and Gallium experiments the deviations are very
large. Notice, in particular, that in several cases only a very
tiny suppression is obtained for the Gallium experiments, though
for the fits to the spectrum with the best g.o.f. -- fits 3a and
4c (see Table 4) -- this is not the case. This underscores a
basic incompatibility, within the VEP picture, of the results from the
different experiments.

In Fig. 2 we show the 90\% C.L. {\em allowed} region  in the
$\sin^22\theta$--$\Delta F$ plane obtained by fitting the
scattered electron spectrum for the $X_B = X_h = 1$ case (area
{\em enclosed} by the solid lines). The best-fit point is marked
with a dark dot. Also shown is the 90\% C.L. {\em disallowed}
region when $X_B$ is permitted to float arbitrarily but keeping
$X_h$ fixed at unity (area {\em enclosed} by the broken lines).
The best-fit point in this case has been indicated by a $\times$
sign. Both (a) active and (b) sterile neutrino alternatives have
been displayed.

\section {Combined analysis of rates and spectrum in the VEP scenario}

In the previous sections we have examined the solar neutrino total
rates and the SK scattered electron energy spectrum within the VEP
oscillation framework.  For the latter, we found a good fit using SSM
input parameters and even better ones when the absolute normalizations
of the $^8$B and/or $hep$ fluxes are allowed to vary (see Table 4). The
rates fit, on the other hand, was not satisfactory and improved
somewhat if the results from the Chlorine experiment was left out from
the $\chi ^2$ analysis. In this section we make a combined fit to the
rates and spectrum data.

For the combined analysis of rate and spectrum, we take the rate
measurements from the Chlorine experiment, the average of the
Gallex and SAGE results, and the SK 1117-day rate data. To obtain
the $\chi^2$ corresponding to any value of the input parameters,
separate values of $\chi^2$ calculated for the total rates and
the spectrum data are added and then the total is minimized.  The
best-fit values  for VEP oscillations to active and sterile
neutrinos are given in Table 6. The number of degrees of freedom
for this case is (18+3 -- 4)=17. Notice that good fits (cases 5a
and 6a) can be obtained only when both $X_B$ and $X_h$ are
allowed to assume values different from their SSM requirements.

\begin{center}
\begin{tabular}{|c|c|c|c|c|c|c|c|}
\hline
Neutrino&Set&$\sin^2 2\theta$ & $\Delta F$ & $X_B$ & $X_h$ &
$\chi^2_{min}$/d.o.f. & g.o.f \\
Type&&& $(10^{-24})$ & & & & (\%)\\
\hline
&5a&1.0 & 1.72 & 0.79 & 23.69 & 11.51/17&82.87 \\ \cline{2-8}
&5b&1.0 & 1.68 & 0.79 & 1.0 & 12.96/18&79.38 \\
&&&&& (fixed) && \\ \cline{2-8}
Active&5c&1.0 & 19.26 & 1.0 & 33.57 & 50.79/15&8.9 $\times 10^{-4}$ \\
&&&&(fixed) &&& \\ \cline{2-8}
&5d&1.0 & 18.75 & 1.0 & 1.0 & 70.44/16&8.37 $\times 10^{-7}$ \\
&&&&(fixed) & (fixed) && \\ \hline
\hline
&6a&1.0 & 1.77 & 0.91 & 34.55 & 14.04/17&66.45 \\ \cline{2-8}
&6b&1.0 & 1.86 & 0.91 & 1.0 & 16.28/18&57.28  \\
&&&&& (fixed) && \\ \cline{2-8}
Sterile&6c&1.0 & 1.86 & 1.0 & 36.87 & 26.90/15& 2.95 \\
&&&&(fixed) &&& \\ \cline{2-8}
&6d&1.0 & 1.89 & 1.0 & 1.0 & 29.84/16&1.88 \\
&&&&(fixed) & (fixed) && \\
\hline
\end{tabular}
\end{center}
\begin{description}
\item{\small \sf Table 6:} {\small \sf The best-fit values of the
parameters, $\sin^22\theta$, $\Delta F$, $X_B$, $X_h$, $\chi^2_{min}$,
and the g.o.f. for fits to the total rates as well as the scattered
electron spectrum at SK.}
\end{description}

In exhibiting the 90\% C.L. allowed regions in parameter space
from this combined fit, we present only the results for the $X_h
= 1, X_B$ arbitrary case. As already noted, if the SSM
normalization for the $^8$B flux is used ($X_B = 1$) then the
goodness-of-fits are poor. These cases are not  pursued further.
In Fig. 3 we present the 90\% C.L. allowed regions
in the $\sin^22\theta$ - $\Delta F$ plane from a combined
analysis of total rates and spectrum. The best-fit points are
marked by a $\times$ sign. Both (a) active and (b) sterile 
neutrino alternatives have been considered. It is noteworthy that
the allowed regions have a considerable overlap with that in Fig.
1 obtained from a fit to the total rates alone.

\section{Predictions for SNO}

The analyses described in the previous sections yield sets of
best-fit values for the parameters $\sin^2 2\theta$, $\Delta F$,
$X_B$ and $X_h$. In this section we examine what these best-fit
values imply for the Sudbury Neutrino Observatory (SNO)
experiment \cite{sno}.

At SNO the neutrinos are detected by three processes, namely, (a)
charge current (CC) break up of the deuteron, (b) electron
scattering by the neutrino, and (c) neutral current (NC) break up
of the deuteron.
\begin{eqnarray}
\nu_e + d & \rightarrow & p + p + e^- ~~~~~~~{\rm (CC\;\;
reaction)}, 
\\
\nu + e^- & \rightarrow & \nu + e^- ~~~~~~~{\rm (scattering)}
\label{scatSNO}, \\
\nu + d & \rightarrow & \nu + p + n ~~~~~~~ {\rm (NC\;\;
reaction)}. \label{ncSNO}
\end{eqnarray}
For the scattering (\ref{scatSNO}) and NC (\ref{ncSNO})
reactions, $\nu$ stands for any active neutrino; it has to be
borne in mind that for the $\mu$ and $\tau$ flavours only the $Z$
exchange contribution is present for the former reaction while
for the $\nu_e$ there is an additional (dominant) piece from
$W$-exchange. For the CC reaction and for scattering, the
electrons are detected by the emitted \u{C}erenkov radiation and
hence their energy spectrum is directly measured as at SK. For the
NC reaction, on the other hand,  only calorimetric measurment is
possible. At present, data is being taken for the first two
processes.

For the scattered electrons, the formalism for theoretical
prediction is similar to that for SK (see eq. (\ref{rkam}))
except for the fact that here in SNO, the detector fluid is 1K
ton of $D_2O$ instead of 32K ton of water in SK.

For the CC interactions, the theoretical predictions for
the electron energy spectrum at SNO is given by
\begin{equation}
T_{SNO}^{th} = 
\sum_\alpha\int_{E_{\nu_{min}}}^{E_{\nu_{max}}} dE_\nu
\int dE_{T} \int_{E_{A_{th}}} dE_{A}
\;\rho(E_{A}, E_{T}) \;X_\alpha \;\phi _\alpha(E_\nu)\;
<P_{ee}(E_\nu)>_\alpha \;\sigma^{CC}.  
\end{equation}
The charge current scattering cross-section at SNO, 
$\sigma^{CC}$, is found from Ref. \cite{snocr}.
In our calculations, we use the resolution function $\rho(E_{A},
E_{T})$ for SNO as given in \cite{blk}. We have set $E_{A_{th}}$
to 5.0 MeV.

In Fig. 4 we show the signal expected at SNO due to (a) $\nu - e$
scattering and (b) the CC reaction if the VEP oscillation is
operative. We choose two typical cases for the VEP parameters:
(i) the best-fit values obtained from the fit to the spectrum using
the SSM inputs ($X_B = X_h = 1$), and (ii) the best-fit parameters
from the combined rate and spectrum fit when both $X_B$ and $X_h$
are allowed to vary. The SSM expectation in the absence of VEP
transitions is also shown. 

Figs 5(a) and 5(b) are similar to Figs. 4(a) and 4(b) excepting
that we present the expectations at SNO normalized by the SSM
predictions. This form may be of more convenience for comparison
with the experimental results.

SNO will subsequently also detect neutrinos {\em via} the NC
reaction (\ref{ncSNO}). Since all active neutrinos register in the
NC reaction with the same strength, this reaction provides a means to
distinguish oscillations to active neutrinos from those to
sterile ones. The ratio of the NC and CC rates, $R_{NC}$ and
$R_{CC}$, is somewhat less
sensitive to theoretical uncertainties than the rates themselves.
Therefore, for the purpose of illustration, we present in Table 7
the values for the ratios
\begin{eqnarray}
R_{SNO} &=& \frac{R_{NC}}{R_{CC}},\\ \nonumber 
S_{SNO}&=& R_{SNO}/R^{BP98}_{SNO} = \frac{R_{NC}/R_{NC}^{BP98}}
{R_{CC}/R_{CC}^{BP98}}\;\;.
\end{eqnarray}
For the NC cross-sections we use Ref.
\cite{snocr}. Results are presented for the best-fit values of the
parameters obtained from the total rates, the SK scattered electron
spectrum, and the combined rates and spectrum data presented in
Tables 2, 4, and 6.

It is seen from Table 7 that the variables $R_{SNO}$ and
$S_{SNO}$ are particularly effective in distinguishing VEP
oscillations to active neutrinos from those to a sterile
neutrino.  For active neutrinos $R_{SNO}$ ($S_{SNO}$) varies
between 0.44 -- 0.74 (1.15 -- 1.92)  while the corresponding
range for sterile neutrinos is 0.29 -- 0.30 (0.77 -- 0.79).

\begin{center}
\begin{tabular}{|c|c|c|c|c|c|c|c|}
\hline
Neutrino&Set&$\sin^2 2\theta$ & $\Delta F$ & $X_B$ & $X_h$ &
$R_{SNO}$&$S_{SNO}$\\
Type&&& $(10^{-24})$ & & & & \\
\hline
&1a,c,d,e&1.0 & 1.80 & 1.0 & 1.00 & 0.610&1.57 \\
\cline{2-8}
&1b&0.85 & 4.57 & 1.0 & 1.00 & 0.740&1.92 \\
\cline{2-8}
&3a&1.0 & $1.97\times10^2$ & 0.79 & 15.07 & 0.581&1.50 \\
\cline{2-8}
&3b&0.38 & 0.22 & 0.65 & 1.0  & 0.443&1.15 \\
Active&&&&& (fixed) && \\ \cline{2-8}
&3c&0.69 & 0.23 & 1.0 & -2.20 & 0.724&1.87 \\
&&&&(fixed) &&& \\ \cline{2-8}
&3d&0.68 & 0.23 & 1.0 & 1.0 & 0.712&1.89 \\
&&&&(fixed) & (fixed) && \\ \cline{2-8}
&5a&1.0 & 1.72 & 0.79 & 23.69 & 0.583&1.51 \\
\hline
\hline
&2a,c,d,e&1.0 & 1.84 & 1.0 & 1.00 & 0.305&0.789 \\
\cline{2-8}
&2b&0.84 & 4.53 & 1.0 & 1.00 & 0.303&0.784 \\
\cline{2-8}
&4a&1.0 & 39.33 & 1.03 & 11.52 & 0.300&0.776 \\ \cline{2-8}
&4b&0.34 & 0.22 & 0.67 & 1.0 & 0.299&0.774 \\
Sterile&&&&& (fixed) && \\ \cline{2-8}
&4c&0.98 & 39.33 & 1.0 & 11.94 & 0.299&0.775 \\
&&&&(fixed) &&& \\ \cline{2-8}
&4d&0.58 & 0.23  & 1.0 & 1.0 & 0.302&0.781 \\
&&&&(fixed) & (fixed) && \\\cline{2-8}
&6a&1.0 & 1.77 & 0.91 & 34.55 & 0.302&0.781 \\ 
\hline
\end{tabular}
\end{center}

\begin{description}
\item{\small \sf Table 7:} {\small \sf The ratio of the NC rate
to the CC rate at SNO, $R_{SNO}$, and the same ratio normalized
to the BP98 SSM, $S_{SNO}$ for the best-fit values of the
parameters, $\sin^22\theta$, $\Delta F$, $X_B$, $X_h$ obtained
from fits to the total rates and the SK scattered electron
spectrum. The sets 1 and 2 are from Table 2, 3 and 4 from Table
4, and 5 and 6 from Table 6.}
\end{description}

\section{Discussions and Conclusions}

In this work we have made a detailed examination of the viability
of the VEP oscillation mechanism in the light of the solar
neutrino data.  The parameters in the VEP formalism are $\Delta
F$, a measure of the violation of the weak equivalence
principle, and $\theta$, the mixing angle relating the flavor
basis of neutrinos to the gravitational basis. The data we have
included in the analysis come from the radiochemical Chlorine
experiment, the similar Gallium-based Gallex and SAGE
collaborations, and the \u{C}erenkov technique reliant Kamiokande
and SuperKamiokande (1117-day data) facilities. We have found
that good fits ($\chi ^2$/d.o.f. = 11.66/16 for active neutrinos)
can be obtained to the scattered electron energy spectrum seen at
SK within the VEP mechanism using SSM inputs. The fits are even
better in models in which the absolute normalizations of the $^8$B
and/or $hep$ fluxes are allowed to vary. The preferred oscillation
wavelength is comparable to the earth-sun distance and is reminiscent
of the mass-mixing vacuum oscillation solution. In contrast to the fit
to the spectral data, the fit to the observed rates is poor but
improves somewhat ($\chi ^2$/d.o.f. = 2.04/2 for active neutrinos) if
the Chlorine experiment is left out of the fit. A combined fit to the
rates and spectrum has also been performed and it is found that within
the SSM the goodness is not satisfactory.  However, in a model in which
the absolute normalizations of the $^8$B and $hep$ fluxes are allowed
to vary, the fit is much improved ($\chi ^2$/d.o.f. = 11.51/17 for
active neutrinos with $X_B = 0.79$ and $X_h = 23.69$). The best-fit
values of $\Delta F$ (= 1.7 $\times 10^{-24}$)  and $\sin ^2 2\theta$ (=
1) are closer to those obtained from the fits to the rates alone. We
find that the 90\% C.L. allowed regions are broadly similar whether VEP
transitions occur to active or sterile neutrinos.

It is of interest to compare the results that we have obtained
with those of other recent analyses of the solar neutrino problem
in the VEP picture \cite{gago,casini}. For the fit to the total
rates, our results agree almost exactly with those of
\cite{casini} and are within 5\% of those of \cite{gago}. The
earlier analyses used the 825-day SK results and we must therefore
conclude that the newer data do not make a significant impact
on this fit. In contrast, for the fits to the SK recoil electron
energy spectrum, our best-fit values of $\Delta F$ and $\sin
^22\theta$ are both somewhat smaller than those of
\cite{gago,casini}. We have checked that if we use the 825-day SK
data there is better agreement. We attribute the
difference to the reduction of the higher energy events in the
newer data. For the combined fit to the rate and spectrum, we are
in good agreement with \cite{gago}; the goodness-of-fit has
improved with the newer data.

During the passage of the neutrinos from their point of
production to the solar surface, interactions with the ambient
matter, responsible for the MSW effect, become important. Apart
from a neutral current contribution which affects the masses of
all active neutrino species identically (and is therefore
irrelevant for this discussion), there is a contribution to the
electron neutrino mass $m_{MSW} \simeq \sqrt{2} G_F n_e(r)$,
$G_F$ being the Fermi coupling and $n_e(r)$ the number density of
electrons at a distance $r$ from the center of the sun, due to
charge current interactions. In the presence of VEP and this MSW
contribution, the effective neutrino mass matrix in flavor space
takes the form
\begin{equation}
M =\frac{1}{2} \left| \begin{array}{cc}
E_\nu\Delta F \cos 2\theta  - 2\sqrt{2} G_F n_e(r) &  E_\nu\Delta F
\sin 2\theta  \\
E_\nu\Delta F \sin 2\theta E_\nu & - E_\nu\Delta F \cos 2\theta 
\end{array} \right|,
\label{eq:mmat}
\end{equation}
where we have dropped an irrelevant part proportional to the
identity matrix. The MSW contribution in (\ref{eq:mmat}) inside
the sun turns out to be several orders of magnitude larger than
the terms due to VEP that we have discussed in this work. Recall
that we have assumed the neutrinos to be massless. For the
maximal mixing case ({\em i.e.}, $\sin2\theta =1$), which we have
found for the best-fit solutions, there is no resonance effect
and, in fact, till such time that the neutrino emerges from the
sun, the MSW contribution controls the  masses in
(\ref{eq:mmat}). Inside the sun, the $\nu_e$ is, therefore,  a
mass eigenstate to a very good approximation. The effect of VEP
oscillations begins to manifest itself only from then onwards.
For oscillation to a sterile neutrino, the neutral current
contribution to the $\nu_e$ mass also becomes relevant.  It is of
the same order as the charged current piece and the resultant
effect in this case is much the same as that for the active
neutrino alternative. There is a different region in the
$\sin^22\theta, \; \Delta F$ parameter space (larger $\Delta F$,
non-maximal mixing) where other solutions have been found with
the MSW effect playing an important role
\cite{bahkra,kuo}\footnote{It has, however, been recently
demonstrated that these solutions are inconsistent with the
results from the non-observation of oscillations in accelerator
neutrino experiments \cite{pant}.}.

SuperKamiokande has now obtained the recoil electron spectra
separately for day and night runs using their 1117-day
data \cite{suzukidn}. The oscillation wavelengths corresponding to
the best-fit VEP parameters that we have found are comparable to
the earth-sun distance (1.49 $\times 10^{11}$m). The extra
distance travelled by neutrinos through the earth during the
night runs is insignificant compared to this. This distance can
be readily estimated.  Using the latitude of SuperKamiokande
(36.4$^o$) and the obliquity of the ecliptic (23.5$^o$), we
estimate the {\em maximum} extra distance travelled during the
night run to be 6.40 $\times 10^6$m. To confirm the expectation
that fits to the separate day and night spectra give results
which are not much different, we performed a
$\chi^2$-minimization on this data for oscillation to an active
neutrino.  When $X_B$ and $X_h$ were held fixed at their SSM
values of unity, the best fit corresponds to $\sin^22\theta =
0.71$ and $\Delta F = 0.21 \times 10^{-24}$. For comparison, the
corresponding values from Table 4, set 3d are $\sin^22\theta =
0.68$ and $\Delta F = 0.23 \times 10^{-24}$. If $X_B$ and $X_h$
are allowed to vary then we get $\sin^22\theta = 1.0$ (1.0),
$\Delta F = 1.97 \times 10^{-22}$ ($1.97 \times 10^{-22}$), $X_B
= 0.78$ (0.79), and $X_h = 0.74
\times 10^{-2}$ (15.07). In the parantheses we have given the
values presented earlier in Table 4, set 3a. Notice that 
there is no significant change barring the very small value of
$X_h$. The $X_h$ dependence of $\chi^2$ is slow and this
indicates that the day-night scattered electron spectra is
consistent with no $hep$ neutrinos.

The VEP oscillation wavelengths favored by the data are
comparable to the distance of the sun to the earth. Therefore,
seasonal effects are to be expected if this mechanism is responsible
for the solar neutrino deficit. However, the data from SK is
still not of very high statistics and we have left this analysis
to a future work.

\vskip 30pt
\parindent 0pt

{\large{\bf {Acknowledgements}}}\\

The authors are grateful to the Calcutta University Computer
Centre for the use of their Origin 2000 computer. D.M. and A.R.
were partially supported by the Eastern Centre for Research in
Astrophysics, India. A.R. also acknowledges a research grant from
CSIR, India. A.S. was supported by a research grant from BRNS,
India and now enjoys a fellowship from UGC, India.

\newpage

\begin{center}
{\Large {\bf Figure Captions}}
\end{center}

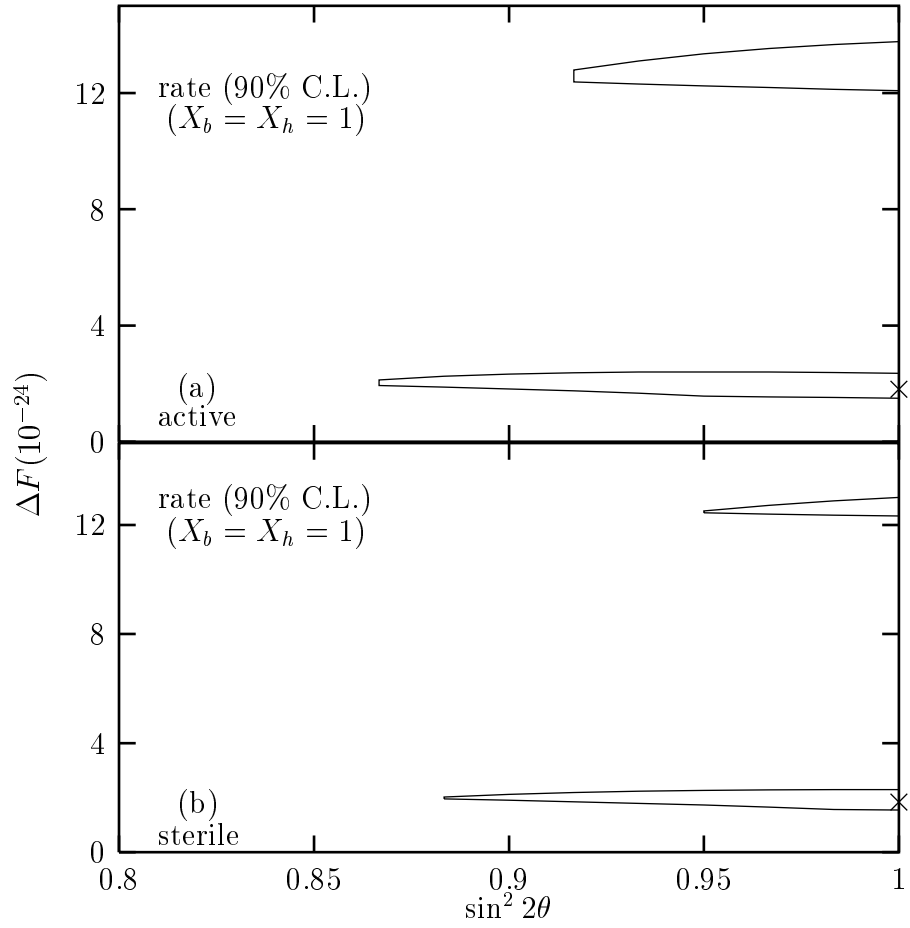
Fig. 1. The 90\% C.L. allowed region in the $\sin^2 2\theta$ -
$\Delta F$ plane from an analysis of the total rates seen by the
Chlorine, Gallex, SAGE, Kamiokande, and SuperKamiokande (1117-day
data) detectors assuming conversion due to VEP to (a) active
neutrinos and (b) sterile neutrinos. The best fit points have
been indicated.\\

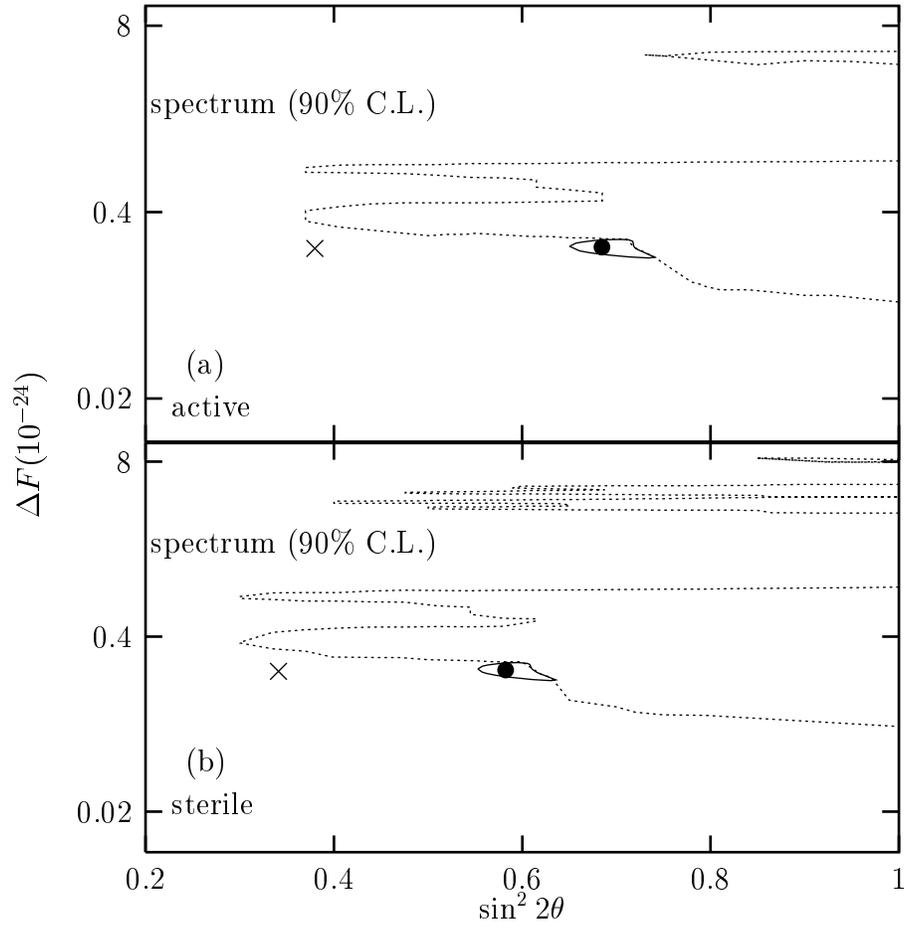
Fig. 2. The region bounded by the solid lines is the 90\% C.L.
{\em allowed} region in the $\sin^2 2\theta$ - $\Delta F$ plane
from the fitting of the 1117-day SK recoil electron spectrum data
with the SSM flux normalizations ($X_B = X_h = 1$). The best fit
point is marked with a dark dot. The best-fit point for the case
$X_h = 1, X_B$ arbitrary, is indicated by a `$\times$' sign and
the corresponding 90\% C.L. {\em disallowed} region is enclosed
by the broken lines. The two panels correspond to (a) active
neutrinos and (b) sterile neutrinos.\\

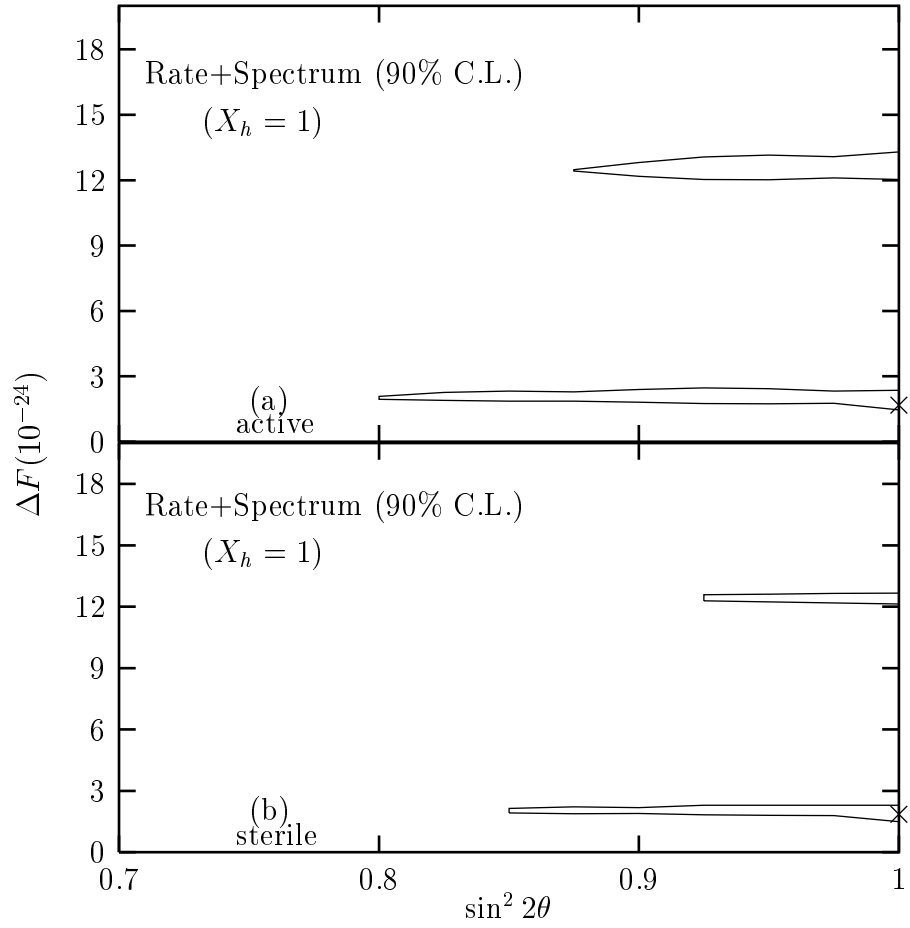
Fig. 3. The 90\% C.L. allowed region in the $\sin^2 2\theta$ -
$\Delta F$ plane from the combined analysis of total rates and
1117-day SK recoil electron spectrum data for conversion due to
VEP. The $hep$ flux normalization  is held fixed at the SSM value
($X_h = 1$) but the $^8$B flux normalization ($X_B$) is allowed
to vary. The best fit points are indicated. Panel (a) represents
the case of conversion to an active neutrino while panel (b)
corresponds to conversion to a sterile neutrino.\\

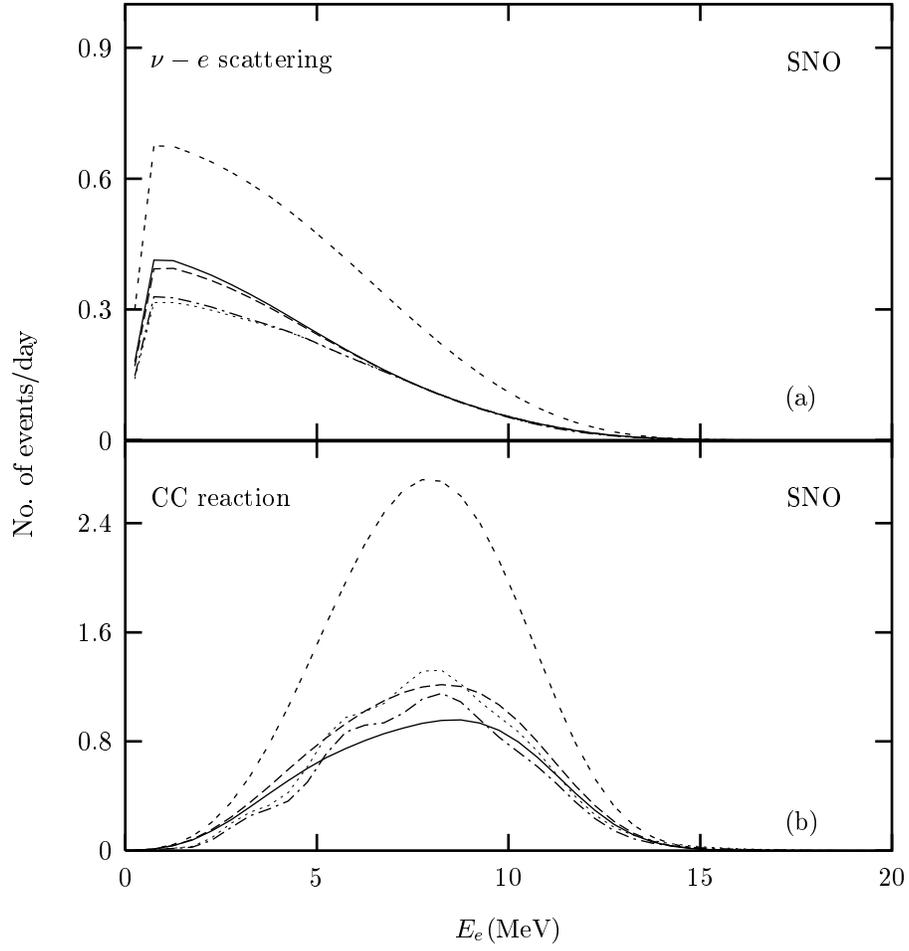
Fig. 4. The number of events expected per day at SNO due to (a)
$\nu - e$ scattering, and (b) the charged current $\nu_e - d$
reaction. Here $E_e$ stands for the observed electron energy.
Results are shown for the best-fit values of the parameters
obtained from fits to the spectrum and jointly to the rates and
the spectrum (see text for more details).  In both panels the
solid (large dashed) line corresponds to the best-fit parameters
obtained from a fit to the electron energy spectrum for VEP
oscillation to an active (sterile) neutrino. The dot-dashed
(dotted) lines similarly correspond to the best combined fits to
the rates and spectrum for VEP oscillations to an active
(sterile) neutrino. The BP98 SSM expectations are the small
dashed curves.\\

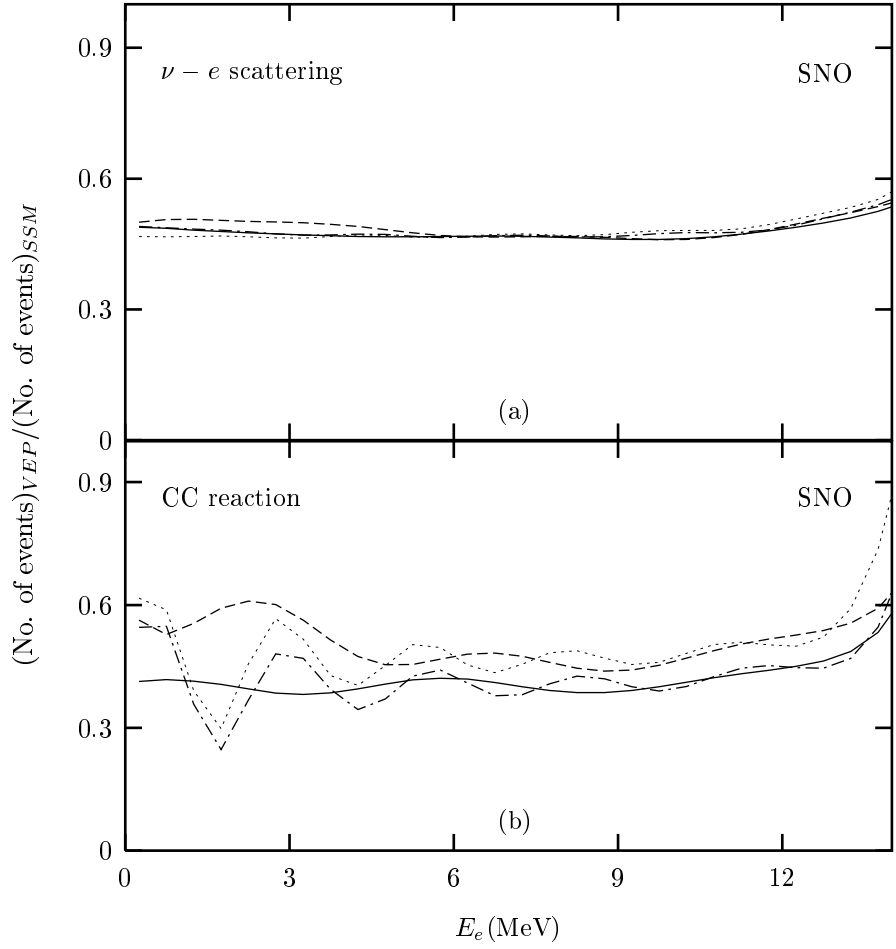
Fig. 5. The ratio of the predicted number of events from VEP to
the SSM expectation as a function of the electron energy for SNO
due to (a) $\nu - e$ scattering, and (b) the charged current
$\nu_e - d$ reaction (see text for more details). Here $E_e$
stands for the observed electron energy. In both panels the solid
(large dashed) line corresponds to the best-fit parameters
obtained from a fit to the electron energy spectrum for VEP
oscillation to an active (sterile) neutrino. The dot-dashed
(dotted) lines similarly correspond to the best combined fits to
the rates and spectrum for VEP oscillations to an active
(sterile) neutrino.\\

\end{document}